\documentclass[aps,pra, twocolumn,nofootinbib, showpacs]{revtex4-1} 
\usepackage[unicode=true,pdfusetitle, bookmarks=true,bookmarksnumbered=false,bookmarksopen=false, breaklinks=false,pdfborder={0 0 0},backref=false,colorlinks=false] {hyperref}
\hypersetup{ colorlinks,linkcolor=myurlcolor,citecolor=myurlcolor,urlcolor=myurlcolor}
\usepackage{braket,colortbl,amsthm,amsmath,cleveref,amssymb,txfonts}\definecolor{myurlcolor}{rgb}{0,0,0.7}
\usepackage{graphics,graphicx}
\usepackage{color}
\usepackage{amssymb}
\usepackage{amsthm}
\usepackage{amsfonts}
\usepackage{float}
\usepackage{graphicx}
\usepackage[format = plain,labelfont = bf,up, textfont = normal , up, justification =raggedright, singlelinecheck =false]{caption}
\usepackage{subcaption}
\usepackage{tabularx}
\usepackage{amsmath}
\usepackage{braket}

\usepackage{graphicx}
\usepackage[utf8x]{inputenc}
\usepackage{color,soul}
\usepackage{amsmath}
\usepackage{braket}
\usepackage{latexsym}
\usepackage{amssymb}
\usepackage{amsthm}
\usepackage{xcolor}
\usepackage{bm}
\usepackage{graphics,epstopdf}
\usepackage{color}\usepackage{amsmath}
\usepackage{enumitem}
\usepackage{fmtcount}
\usepackage{booktabs}
\usepackage{epsfig}
\theoremstyle{plain}

\def\bea{\begin{eqnarray}}
\def\eea{\end{eqnarray}}
\def\ba{\begin{array}}
\def\ea{\end{array}}

\def\ket{\rangle}
\def\bra{\langle}
\def\beq{\begin{equation}}
\def\eeq{\end{equation}}

\begin{document}
\title{Detection loophole in measurement-device-independent entanglement witness}

\author{Kornikar Sen, Chirag Srivastava, Shiladitya Mal, Aditi Sen(De), Ujjwal Sen}

\affiliation{Harish-Chandra Research Institute, HBNI, Chhatnag Road, Jhunsi, Allahabad 211 019, India}

\begin{abstract}
There always exists an entanglement witness for every entangled quantum state. Negativity of the expectation value of an entanglement witness operator guarantees entanglement of the corresponding state, given that the measurement devices involved are perfect, i.e., the performed measurements actually constitute the witness operator for the state under consideration. In a realistic situation, there are two possible ways of measurements to drive the process away from the ideal one. Firstly, wrong measurements may be performed, and secondly, while the measurement operators are implemented correctly, the detection process is noisy. Entanglement witnesses are prone to both of these imperfections. The concept of measurement-device-independent entanglement witnesses was introduced to remove the first problem. We analyze the ``detection loophole'' in the context of measurement-device-independent entanglement witnesses, which deal with the second problem of imprecise measurements.  We obtain an upper bound on the entanglement witness function in the measurement-device-independent entanglement witness scenario, below which entanglement is guaranteed for given non-ideal detector efficiencies, that can involve both lost events and dark counts. 
\end{abstract}
\maketitle

%We subsequently compare it with the bound on the corresponding entanglement witness function in the measurement-device-dependent entanglement witness scenario for the same non-ideal detector efficiency.  

\section{Introduction}
\label{sec1}
 Quantum mechanics provides description for physical systems in terms of quantum states belonging to some Hilbert space. One of the distinguishing features of composite quantum systems is the presence of a kind of correlation between different subsystems, called entanglement \cite{HHHH, das'06, toth'09}. Presence of entanglement enables better efficiencies of various tasks like quantum teleportation \cite{bennett'93}, quantum dense coding \cite{bennett'92}, and entanglement-based quantum cryptography \cite{ekert'91}. Therefore, it is of importance to know whether a state is entangled. Over the years,  several methods have been proposed to detect entanglement, and to name a few, there are the positive partial transpose criterion \cite{peres'96,horodecki'96}, range criterion \cite{horedecki'97}, violation of Bell inequality \cite{bell}, and negativity of the expectation values of entanglement witness (EW) operators \cite{horodecki'96,terhal'00}. 
 Several such procedures have also been used to verify entanglement in laboratories
% 
% See 
\cite{exp1}.
%and references therein for experimental detections of entanglement. 

  The method of entanglement witnesses has gained importance over the years as an efficient detector of entanglement of shared quantum states. If one has some \emph{a priori} information about a shared quantum state, then an EW may be constructed for that state. It is also possible to implement such a method in a laboratory as the expectation value of any EW can be obtained by performing local measurements on subsystems constituting the composite system. Still, an imperfect implementation of the measurements may wrongly indicate a separable state to be entangled \cite{witness1}, just as a ``local" state may appear to violate a Bell inequality \cite{loophole-BI}.
  Negativity of the expectation value of any entanglement  witness operator guarantees entanglement, when the measurement devices involved are ideal. There are at least two possible ways by which realistic measurements can drive the process away from its ideal variety. Firstly, the intended measurement basis may get altered (``wrong" measurements) and secondly, the detectors used can be noisy (``imprecise" measurements). EWs are prone to both these ``defects". Both pose challenges to an experimentalist, as defective measurements can lead to false positives in witnessing entanglement. The concept of measurement-device-independent entanglement witnesses (MDI-EWs) was introduced to address the first problem \cite{Cyril'13}. It was shown that an MDI-EW can always be constructed from a standard EW, based on a semi-quantum nonlocal game, where every entangled state provides advantage over all separable states \cite{Buscemi'12}.
  %To resolve this issue, Branciard \emph{et. al.} \cite{Cyril'13} came up with the idea of measurement-device-independent entanglement witnesses (MDI-EWs). 
Robustness of MDI-EWs as compared to that of standard ones, for sequential witnessing of entanglement was analyzed in \cite{srivastava}.   
  
%In this game, observers sharing a multipartite state perform joint measurements on their part of shared state and a quantum input to constitute a pay-off function. Entangled states return higher values of pay-off function than all the separable states. This semi-quantum nonlocal game differs from usual non-local games (corresponding to some Bell inequality) in the sense that quantum inputs are allowed instead of classical inputs.  \\

 In this article, we focus on the second problem, which involves imprecise measurements due to noisy detectors. Any of the entanglement detection methods, if performed with imprecise measurements, can lead to erroneous observations. By ``imprecise measurements", we mean here that the corresponding measurement devices have non-unit efficiency, arising due to loss of measurement outcomes. Such studies are usually clustered into what is referred to as the ``detection loophole'' problem, as the task is to detect entanglement with non-ideal measurement devices. The problem of detection loophole 
 %in the context of  
 for 
 Bell inequality violations has been studied extensively, both theoretically \cite{loophole-BI} as well as experimentally 
 %. There exist several experimental works on violation  of  Bell  inequalities with non-unit detector efficiencies 
 \cite{BIexp}. 
 %For analysis of the detection loophole problem i
 In the context of entanglement witnesses, the detection loophole problem was analyzed recently in  \cite{witness1, witness2}.

The purpose of MDI-EWs is to remove the measurement-device-dependence of EWs, i.e., to guarantee entanglement of a shared quantum state, independent of what measurements are being performed. But whether this type of witness also guarantees entanglement, independent of the possibly non-unit efficiencies of the corresponding measurement devices, is yet to be answered. In this article, we analyze the detection loophole present in MDI-EWs considering such non-ideal measurement devices. To be specific, we  consider the problem in three scenarios.
%two kinds of detector inefficiencies. 
The first one involves the losses in the detector, whereas the second considers the additional counts (``dark counts") therein. We finally consider a third scenario, where both these kinds of detector inefficiencies can be present.
%\textbf{
In the case of lossy detectors, having a certain non-unit efficiency, we provide an upper bound on the measured value of the MDI-EW function which is sufficient to certify entanglement. 
%A comparison of this upper bound, in case of lossy detectors, is made with the upper bound on the witness function in EW scenario.
In case of additional counts, we observe that MDI-EWs do not guarantee entanglement falsely, whatever be the detector's efficiency. We finally derive a relation for loophole-free detection of entanglement using a MDI-EW in presence of both types of detector inefficiencies. We illustrate the results by using Werner states \cite{Werner} and noisy Greenberger-Horne-Zeilinger (GHZ) states \cite{gange-Dheu-khele-jai}.
%}

%In the case of lossy detectors, it is observed that MDI-EWs do not always guarantee entanglement, and we find the condition for which an MDI-EW do perform better than standard EWs. In case of additional counts, MDI-EWs still guarantee entanglement as in the case of standard EWs.

This paper is organized as follows. In Sec. \ref{sec2}, we discuss about the detection loophole in case of standard EWs. In Sec. \ref{sec3}, a brief description on MDI-EWs is given. The detection loophole in case of MDI-EWs is analyzed in Sec. \ref{sec4}. A short summary is provided in Sec.\ref{sec5}.

 %SIGNIFICANCE OF mdiew USING ew

% Most of these entanglement detection critrion involve measurements. loophole.    Theoretically, one can always choose the measurements to be ideal, i.e., Each method has its own applicability, for example, the PPT criteria provides a sufficient condition for entanglement detection in few cases only, violation of Bell inequality provides sufficient condition but not all entangled states are detected, and an entanglement witness operator can be constructed if some information about the quantum state is known apriori. 

\section{Entanglement witnesses and the detection loophole}
\label{sec2}
In this section, we briefly discuss about EWs and ways of overcoming the detection loophole in the process of measuring expectation values corresponding to an EW \cite{witness1}. 
A hermitian operator, $W$, is called an EW of a certain shared system, if and only if tr$(W\sigma)\geq 0$ for all separable states $\sigma$ of that system and there exists at least one entangled state $\rho_e$ such that tr$(W\rho_e)<0$. It can be seen that for every entangled state, $\rho_e$, there exists at least one witness operator, $W$, which satisfies the above conditions and can detect that $\rho_e$ \cite{horodecki'96}. Let us mention two examples of witness operators. We know that the Werner state \cite{Werner}, $\rho_p=p |\psi^- \ket \bra \psi^- | +\frac{1-p}{4} \mathbb{I}_4$, where $|\psi^- \ket=\frac{1}{\sqrt{2}}(|01 \ket -|10 \ket)$ and $0\leq p \leq 1$, 
%on the Hilbert space $\mathcal{H}_A \otimes \mathcal{H}_B$, 
is entangled for $p > \frac{1}{3}$, with $\{ |0 \rangle, |1\rangle \}$ being the eigenbasis of the Pauli $\sigma_z$ operator, as ensured by the partial transposition criterion \cite{peres'96,horodecki'96}. 
%Here $\mathcal{H}_A$ and $\mathcal{H}_B$ are both qubit Hilbert spaces, and $\{ |0 \rangle, \text{ } |1\rangle \}$ form the eigenbasis of the Pauli $\sigma_z$ operator. 
Here, $\mathbb{I}_k$ denotes the identity operator on a complex Hilbert space of dimension $k$. For $p > \frac{1}{3}$, the operator $\rho_p^{\text{T}_B}$ will have one negative eigenvalue \cite{neg}. Here, the superscript $\text{T}_B$ denotes transposition on the second system.
%Hilbert space $\mathcal{H}_B$.  
The eigenvector corresponding to this negative eigenvalue, viz. $\frac{1}{4} (1 - 3 p)$, is $|\phi^+\ket=\frac{1}{\sqrt{2}} (|00\ket +|11 \ket) $. Then a witness operator that can detect the state $\rho_p$ can be considered to be of the form \cite{horodecki'96, terhal'00}
\begin{equation}
 W_{\rho_p}=| \phi^+ \ket \bra \phi^+ |^{T_B}. \label{W1}
\end{equation} 
 Another example can be given by considering 3-qubit noisy GHZ states, $\rho_q^{GHZ}=q|GHZ \ket \bra GHZ |+ \frac{1-q}{8} \mathbb{I}_8$, where $|GHZ \ket = \frac{1}{\sqrt{2}}(|000 \ket +| 111\ket)$ \cite{gange-Dheu-khele-jai} and $0\leq q\leq 1$. This state 
 can be shown to be
 genuinely tripartite entangled for $q> \frac{3}{7}$ \cite{ref1}. A witness operator which can detect genuine tripartite entanglement of $\rho_q^{GHZ}$ is given by \cite{toth'09,ref2} 
  \begin{equation}
  W_{\rho_q^{GHZ}}=\frac{1}{2}\mathbb{I}_8-|GHZ \ket \bra GHZ |. \label{W2}   
\end{equation}

But, due to existence of the detection loophole in measurement processes, one may mistakenly identify a separable state, say $\sigma'$, as an entangled state, i.e, one may get tr$\left( W\sigma'\right)< 0$, where $W$ is the corresponding entanglement witness. The condition for overcoming the detection loophole for entanglement witnesses was first discussed by Skwara \emph{et al.} \cite{witness1}.
Since we can always decompose a hermitian operator, acting on a tensor product Hilbert space, $\mathcal{H}_1\otimes \mathcal{H}_2 \ldots \otimes \mathcal{H}_n$, in terms of the identity operator, $\mathbb{I}$, of the total Hilbert space and traceless hermitian operators, $S_\alpha$, as products of hermitian operators operating on local subspaces $\mathcal{H}_1$, $\mathcal{H}_2$, $\cdot \cdot \cdot$, $\mathcal{H}_n$, we can write an $n$-partite witness operator as 
\begin{equation}
W=C_0\mathbb{I}+\sum_\alpha C_\alpha S_\alpha \label{eq1}, \end{equation}    
where $C_0$ and $C_\alpha$ are real numbers. Now, an expectation value of $W$ depends on the individual expectation values of $S_\alpha$. The measured expectation value of $S_\alpha$ for a state $\rho$ is given by $\langle S_\alpha \rangle_m=\frac{\sum_i n_i \lambda_i}{N} $, where $n_i$ denotes the number of times the eigenvalue $\lambda_i$ has clicked and $N=\sum_i n_i$. But due to additional and lost events in the measurement process, this value could be different from the true expectation value of $S_\alpha$ given by $\langle S_\alpha\rangle _t=\frac{\sum_i \tilde{n_i} \lambda_i}{\tilde{N}}$. Here $\tilde{n_i}$ denotes number of times $\lambda_i$ should have clicked in the case of ideal detectors and $\tilde{N}=\sum_i \tilde{n_i}$. Let the number of additional and lost events for $\lambda_i$ be $\epsilon_{+i}$ and $\epsilon_{-i}$ respectively, and set $\sum_i \epsilon_{\pm i}=\epsilon_{\pm}$. The additional and lost event efficiencies are defined as $\eta_+=\frac{\tilde{N}}{\tilde{N}+\epsilon_+}$ and $\eta_-=\frac{\tilde{N}-\epsilon_-}{\tilde{N}}$.

In this paper, we consider the following three cases separately. In the first case (\emph{Case 1}), the additional event efficiency (i.e. $\eta_+$) is taken to be unity, with the lost event efficiency (i.e. $\eta_-$) being kept arbitrary. 
%while i
In the next case (\emph{Case 2}), keeping $\eta_-=1$, we consider arbitrary $\eta_+$. Finally, in the third case (\emph{Case 3}), we consider the general case of arbitrary additional and lost event efficiencies.
In all the cases, $\epsilon_{\pm i}$'s are taken to be independent of $i$, say $\overline{\epsilon}_\pm$. 

\emph{Case 1:} For the first case, we get $\langle S_\alpha \rangle _m=\frac{\sum_i \left(\tilde{n}-\overline{\epsilon}_-\right)\lambda_i}{\tilde{N}-\epsilon_-}=\left(\frac{\sum_i \tilde{n_i}\lambda_i}{\tilde{N}}-\frac{\overline{\epsilon}_-\sum_i \lambda_i}{\tilde{N}}\right)\frac{\tilde{N}}{\tilde{N}-\epsilon_-}$. Since $S_k$ are traceless matrices, we get $\langle S_\alpha \rangle _m=\frac{\langle S_\alpha \rangle _t}{\eta_-}$. Hence using Eq. \eqref{eq1}, we get the relation between measured and true values of $W$ as 
\begin{eqnarray}
\langle W\rangle_m&=&C_0+\sum_\alpha C_\alpha \frac{\langle S_\alpha \rangle _t}{\eta_-} \nonumber \\
&=&C_0\left(1-\frac{1}{\eta_-}\right)+\frac{\langle W \rangle_t}{\eta_-}. \nonumber
\end{eqnarray}  
Now, as we mentioned before, $W$ will detect entanglement of a state $\rho$ when $\langle W \rangle_t < 0$, where the expectation value is taken over the state $\rho$. Hence, even in the presence of detection loophole, $W$ can detect the entanglement of $\rho$ correctly when the following inequality is satisfied:
\begin{equation}
\langle W\rangle_m < C_0\left(1-\frac{1}{\eta_-}\right).  \label{eq3}
\end{equation} 
As the completely depolarized state is a separable state, we must have $C_0>0$. Therefore, $C_0\left(1-\frac{1}{\eta_-}\right) < 0$ for $\eta_- \neq 1$. States for which $C_0\left(1-\frac{1}{\eta_-}\right) \leq \langle W \rangle_m < 0$, a negative $\langle W\rangle_m$ will not be sufficient for the experimentalist to infer entanglement in the shared state. 
If we consider the two examples of witness operators that we considered above, viz. the conditions for Werner state ($W_{\rho_p}$) and noisy GHZ state ($W_{\rho_q^{GHZ}}$), we get the conditions
\begin{eqnarray}
\langle W_{\rho_p}\rangle_m < \frac{1}{4}\left(1-\frac{1}{\eta_-}\right)  \text{ and}\nonumber \\ 
\langle W_{\rho_q^{GHZ}}\rangle_m < \frac{3}{8}\left(1-\frac{1}{\eta_-}\right),\nonumber
\end{eqnarray}
for loophole-free detection of entanglement in the corresponding states, in the case when the lost event efficiency is non-ideal.
Hence, for example, if the lost event efficiency is $\eta_-=\frac{1}{2}$, then the witness operator, $W_{\rho_p}$, can detect an entangled Werner state if $\langle W_{\rho_p}\rangle_m < -\frac{1}{4}$ is satisfied,  and $W_{\rho_q^{GHZ}}$ can detect an entangled noisy GHZ state if $\langle W_{\rho_q^{GHZ}}\rangle_m < -\frac{3}{8}$ is satisfied.

\emph{Case 2:} Doing similar calculations for the second case, we get
\begin{equation}
\langle W\rangle_m=C_0\left(1-\eta_+\right)+\eta_+ \langle W \rangle_t. \label{eq4}
\end{equation}
Here, $C_0\left(1-\eta_+\right) \geq 0$. Hence $ \langle W \rangle_t \leq \langle W\rangle_m$ for all values of $\eta_+$. Therefore, in this case, even though lesser states will be detected using $W$, we will never identify a separable state as an entangled state. Hence, EW is robust under the second kind of loophole.

\emph{Case 3:} Let us now move to 
%Finally, let us consider 
the general case where the additional event efficiency, $\eta_+$, as well as the lost event efficiency, $\eta_-$, are both at non-unit  values.
%kept arbitrary. 
The condition on the measured expectation value of witness function, $\bra W \ket_m$, which guarantees entanglement, 
even in the presence of non-ideal \(\eta_{\pm}\),
is given by 
%the relation
\begin{equation}
\bra W \ket_m<C_0\left(1-\frac{1}{\eta_-+\frac{1}{\eta_+}-1}\right).
\label{pawn}
\end{equation}
Notice that \(\eta_-=1\) or \(\eta_+=1\) lead us to the previously obtained relations.

\section{Measurement device independent Entanglement witness}
\label{sec3}

 Any entanglement witness operator can always be expanded in a basis consisting of tensor products of local hermitian operators. Expectation value of a witness operator in any given quantum state can therefore be obtained by performing measurements in the bases of those local hermitian operators. However, if an experimentalist performs a measurement that is different from what is required to evaluate the expectation value of the witness operator in a given state, then a negative expectation value does not guarantee that the state is entangled.  
   Therefore, Branciard \emph{et al.} \cite{Cyril'13} came up with the idea of measurement-device-independent entanglement witnesses (MDI-EWs) which guarantee the entanglement of a quantum state, i.e., it never indicates a separable state to be entangled, independent of the measurements that are actually performed. Moreover, it was also observed that when effect of lossy detectors is to modify joint probabilities by the same multiplicative factor, then MDI-EWs can still do its job properly \cite{Cyril'13}. 
   
   In this section, we briefly discuss the MDI-EWs studied in \cite{Cyril'13}, motivated by the ``nonlocal semi-quantum game'' presented in \cite{Buscemi'12}. Let $W$ be an entanglement witness operator acting on the Hilbert space $\mathcal{H}_A \otimes \mathcal{H}_B$, with the dimension of $\mathcal{H}_A$ being $d_A$ and that of $\mathcal{H}_B$ being $d_B$. Consider complete sets of density matrices, $\{\tau_r|~r=1,2,\ldots,d_A^2\}$ acting on $\mathcal{H}_A$ and $\{\omega_s|~s=1,2,\ldots,d_B^2\}$ acting on $\mathcal{H}_B$. These sets are ``complete" in the sense that they span the space of hermitian matrices on the corresponding Hilbert spaces ($\mathcal{H}_A$ or $\mathcal{H}_B$) with respect to the field of reals. One can always find a set of real numbers, ${\beta_{rs}}$, such that any entanglement witness can be expanded as
\begin{equation}
\label{witness}
W=\sum_{rs}\beta_{rs}\tau_r^ \text{T} \otimes \omega_s^\text{T},  
\end{equation}
where the superscript, T, over the states denotes their transposes.
 The above equation is a consequence of the existence of a set of density matrices which span the space of hermitian operators. Note that the expansion in Eq. (\ref{witness}) is not unique.
   
Consider two parties Alice $(A)$ and Bob $(B)$ sharing a quantum state $\rho_{AB}$ acting on $\mathcal{H}_A \otimes \mathcal{H}_B$. Let $A$ and $B$ be provided the set of states $\{ \tau_r\}$ and $\{ \omega_s\}$ as their auxiliary quantum inputs, respectively.

Each party then performs a joint dichotomic measurement, \(\{|\Phi_{A(B)}^{+}\ket\bra\Phi_{A(B)}^{+}|,~\mathbb{I}_{d_{A(B)}^2}-|\Phi_{A(B)}^{+}\ket\bra\Phi_{A(B)}^{+}|\}\), on their part of the shared state $\rho_{AB}$ and their respective inputs. Here $|\Phi_{A(B)}^{+}\ket=\frac{1}{\sqrt{d_{A(B)}}}\sum_{i=0}^{d_{A(B)}-1}|ii\ket$, whereas $\{|i\ket\}$ forms the computational basis. For simplicity of notations, the outcomes $|\Phi_{A(B)}^{+}\ket\bra\Phi_{A(B)}^{+}|$ will be indicated as `1', and $\mathbb{I}_{d_{A(B)}^2}-|\Phi_{A(B)}^{+}\ket\bra\Phi_{A(B)}^{+}|$ will be indicated as `0'. The joint probability of $A$ getting an outcome $a$ given $\tau_r$ as the quantum input, and $B$ getting an outcome $b$ given $\omega_s$ as the quantum input is denoted by $P(a,b|\tau_r, \omega_s)$. 
The MDI-EW function is then given by 
\begin{equation}
\label{I}
I(\rho_{AB})=\sum_{r,s} \beta_{rs}P (1,1|\tau_r, \omega_s).
\end{equation}
The connection between the MDI-EW function, $I(\rho_{AB})$, and the entanglement witness, $W$, can be obtained with the help of Eq. (\ref{witness}), and is given by \cite{Cyril'13}
\begin{equation}
\label{witness_I}
I(\rho_{AB})=\frac{\text{tr}(W\rho_{AB})}{d_Ad_B},
\end{equation}
where $\rho_{AB}$ is the shared state whose entanglement detection is in process. The MDI-EW function thus satisfies all the properties of an entanglement witness operator.

Suppose now that $A$ and $B$ share a separable state and let instead of performing the POVM, \(\{|\Phi_{A(B)}^{+}\ket\bra\Phi_{A(B)}^{+}|,~\mathbb{I}_{d_{A(B)}^2}-|\Phi_{A(B)}^{+}\ket\bra\Phi_{A(B)}^{+}|\}\), $A$ performs an \emph{arbitrary} two-outcome POVM, \(\{A_0,~A_1\}\), and $B$ performs an \emph{arbitrary} two-outcome POVM, \(\{B_0,~B_1\}\), to construct the MDI-EW function for that separable state. It turns out that the MDI-EW function for any separable state still remains positive, i.e., $I(\sigma_{AB})\geqslant 0$, for all $\sigma_{AB}$ belonging to the space of separable states, thus justifying the ``measurement-device-independent" adjective of MDI-EWs. 

To formulate a MDI-EW for detecting genuine multipartite entanglement (GME), one can use the following decomposition of a witness operator,
\begin{equation}
W_{\rho_q^{GME}}=\sum_{r,s,u}\beta_{rsu}\tau_r^\text{T} \otimes \omega_s^\text{T} \otimes \gamma_u^\text{T}, \label{W5}
\end{equation}
where $W_{\rho_q^{GME}}$ is known to detect the GME of a three-party state $\rho_{GME}$. In this case, the three parties, say Alice, Bob, and Charlie, share the state $\rho_{GME}$ and have $\tau_r$, $\omega_s$, $\gamma_u$ as their auxiliary quantum inputs, and the witness operator acts on the joint Hilbert space of Alice, Bob, and Charlie, which are, say $\mathcal{H}_A$, $\mathcal{H}_B$, and $\mathcal{H}_C$ respectively. Hence, the corresponding MDI-EW is given by \cite{Cyril'13}
\begin{equation}
I(\rho_{GME})=\sum_{r,s,u}\beta_{rsu}P(1,1,1|\tau_r, \omega_s, \gamma_u). \label{W6}
\end{equation}

In the bipartite case, the witness operator, given in Eq. \eqref{W1}, which can detect entanglement of the Werner state, can be decomposed in the form of Eq. \eqref{witness}, where $\beta_{rs}$, $\tau_r$, and $\omega_s$ are given by \cite{Cyril'13}
  \begin{eqnarray}
\beta_{rs}=\frac{5}{8} \text{ for $r=s$, \ \ \ \ \ \ \ \ \ } \beta_{rs}=-\frac{1}{8} \text{ for $r\neq s$}, \nonumber \\
\tau_r=\sigma_r \frac{\mathbb{I}_2+\vec{n} \cdot \vec{\sigma}}{2} \sigma_r, \text{ \ \ \ \ \ \ \ \ \ } \omega_s=\sigma_s \frac{\mathbb{I}_2+\vec{n}\cdot \vec{\sigma}}{2} \sigma_s. \label{W4}
 \end{eqnarray}
Here $r$ and $s$ run from 0 to 3, $\vec{n}=\frac{1}{\sqrt{3}}(1,1,1)$, and $\vec{\sigma}=(\sigma_1,\sigma_2, \sigma_3)$, i.e., the Pauli spin matrices  and $\sigma_0=\mathbb{I}_2$. Hence, the corresponding MDI-EW can be constructed as \cite{Cyril'13}
\begin{equation}
I(\rho_p)=\frac{5}{8} \sum_{s=t} P(1,1|\tau_r, \omega_s)-\frac{1}{8} \sum_{s\neq t} P(1,1|\tau_r, \omega_s). \nonumber
\end{equation}
Similarly, the witness operator for the noisy GHZ state, given in Eq. \eqref{W2}, can be decomposed as in Eq. \eqref{W5}. We chose the same $\tau_r$ and $\omega_s$ as in Eq. \eqref{W4} and determined the corresponding $\gamma_u$. In this particular decomposition, the coefficients $\beta_{r,s,u}$ are given by \cite{Cyril'13}
\begin{eqnarray}
\beta_{rsu}=\frac{3}{32} && \left( -1 \right)^{\left[(r-1)/2 \right]\left[(s-1)/2 \right]+\left[(r-1)/2 \right]\left[(u-1)/2\right]+\left[(s-1)/2 \right]\left[(u-1)/2\right]+1} \nonumber \\
\times && \left(  -1 \right)^{\left[ (r-1)/2\right]+\left[ (s-1)/2\right]+\left[ (u-1)/2\right]}+\left(-1 \right)^{r+s+u}\sqrt{3}. \nonumber
\end{eqnarray}
Using this $\beta_{rsu}$ in Eq. \eqref{W6}, the MDI-EW can be determined and used in experiments.  
\section{DETECTION LOOPHOLE IN MDI-EW}
\label{sec4}

%Although the value of MDI-EW function, independent of the POVM performed on the joint state $\tau_s \otimes \sigma_{AB} \otimes \omega_t$, is always non-negative for all separable states, is this also robust in the presence of detection loophole, i.e, even if the efficiency defined in Sec. \ref{sec2}, $\eta_\pm \neq 0$, will $I$ still satisfies $I(\sigma_{AB}) \geq 0$ for all separable states $\sigma_{AB}$?   

As discussed in the preceding section, the value of an MDI-EW, independent of the POVM performed on the joint state, $\tau_r \otimes \sigma_{AB} \otimes \omega_s$, is always non-negative for all separable states, $\sigma_{AB}$. Now the question is whether it is also robust in the presence of detection inefficiencies. In other words, we ask: \emph{even if the detection efficiencies, defined in Sec. \ref{sec2}, are not unit, will $I(\sigma_{AB}) \geq 0$ hold for all separable states?}  In this section, we explore this question, find that a detection loophole is, in principle, also present in this measurement-device-independent scenario, and derive the condition for closing the detection loophole. 
In the first case (\emph{Case I}), we  take the additional event efficiency, say $\Xi_{\eta_+}$, to be 1 and consider an arbitrary lost event efficiency, say $\Xi_{\eta_-}$. In the second case (\emph{Case II}), $\Xi_{\eta_-}$ is taken to be ideal while 
%will be fixed at 1 and 
$\Xi_{\eta_+}$ can take any value between 0 to 1. 
In the final case (\emph{Case III}), we take up the general situation of arbitrary values for the two event efficiencies.

\emph{Case I:} For the first case, the measured value of probability $P (a,b|\tau_r,\omega_s)$ is given by
\begin{equation}
P (a,b|\tau_r,\omega_s)_m=\frac{n_{ab}^{rs}}{N^{rs}}, \nonumber
\end{equation}
where $n_{ab}^{rs}$ denotes the number of times Alice and Bob respectively got outcomes $a$ and $b$ when the quantum inputs were $\tau_r$ and $\omega_s$, and $N^{rs}=\sum_{a,b}n_{ab}^{rs}$. We assume $N^{rs}=N$ for all $r$ and $s$. From now on, we denote $P (a,b|\tau_r,\omega_s)$ as $P^{ab}_{rs}$, and the measured and true values of $P^{ab}_{rs}$ as $\left( P_{ab}^{rs}\right)_m$ and $\left( P_{ab}^{rs}\right)_t$ respectively. Now, $\left( P_{ab}^{rs}\right)_m$ could be different from the true value of $P_{ab}^{rs}$, given by
\begin{equation}
\left( P_{ab}^{rs}\right)_t=\frac{\tilde{n}_{ab}^{rs}}{\tilde{N}^{rs}}, \nonumber
\end{equation}   
where $\tilde{n}_{ab}^{rs}$ denotes the number of times outcomes $a$ and $b$ would click in the ideal case of no lost events. And $\tilde{N}^{rs}=\sum_{a,b} \tilde{n}_{ab}^{rs}$. We assume that $\tilde{N}^{rs}=\tilde{N}$ for all $r$ and $s$. Now, $n_{ab}^{rs}=\tilde{n}_{ab}^{rs}-\epsilon_{-ab}^{rs}$, where $\epsilon_{-ab}^{rs}$ is the number of lost events with outcomes $a$ and $b$.
For simplicity, we assume that the $\epsilon_{-ab}^{rs}$ are equal for all $a, b, \tau_r,$ and $ \omega_s$, and set $\sum_{a,b}\epsilon_{-ab}^{rs}=\Xi_{\epsilon_-}$. We define the lost event efficiency as $\Xi_{\eta_-}=\frac{\tilde{N}-\Xi_{\epsilon_-}}{\tilde{N}}$. Hence we get the relation between measured and true values of the relevant probability as
\begin{eqnarray}
\left( P_{11}^{rs}\right)_m &=& \frac{\tilde{n}_{11}^{rs}-\frac{\Xi_{\epsilon_-}}{4}}{\tilde{N}-{\Xi_{\epsilon_-}}} \nonumber\\
&=&\frac{\tilde{N}\left(P_{11}^{rs} \right)_t-\frac{\Xi_{\epsilon_-}}{4}}{\tilde{N}-\Xi_{\epsilon_-}} \nonumber\\
&=&\frac{\left( P_{11}^{rs}\right) _t}{\Xi_{\eta_-}}-\frac{1-\Xi_{\eta_-}}{4\Xi_{\eta_-}}. \nonumber
\end{eqnarray}
The factors \(1/4\) appearing in the preceding calculation 
%in the denominator of the RHS of the inequality 
are 
%coming 
due to the fact that $a,b$ can take 4 different values, viz., 00, 01, 10, 11.
We therefore have that the measured value of the MDI-EW is related to its true value via the relation,
\begin{eqnarray}
I_m(\rho_{AB})&=&\sum_{r,s}\beta_{rs}\left ( \frac{\left( P_{11}^{rs}\right)_t}{\Xi_{\eta_-}}-\frac{1-\Xi_{\eta_-}}{4\Xi_{\eta_-}}\right) \nonumber\\
&=& \frac{I_t(\rho_{AB})}{\Xi_{\eta_-}}-\frac{1-\Xi_{\eta_-}}{4\Xi_{\eta_-}}\sum_{r,s} \beta_{rs} . \label{eq2}
\end{eqnarray} 
Now the entanglement in the state, ${\rho}_{AB}$, will be detected if $I_t(\rho_{AB}) < 0$ is satisfied. 
%Putting this in 
From Eq. \eqref{eq2}, 
this implies that 
%we get
\begin{equation}
I_m(\rho_{AB}) < - \frac{1-\Xi_{\eta_-}}{4\Xi_{\eta_-}}\sum_{r,s} \beta_{rs}. \nonumber
\end{equation} 
Since $\sum_{r,s} \beta_{rs}=\text{tr} (W)$, we get
\begin{equation}
I_m(\rho_{AB}) < \frac{\text{tr} (W)}{4}\left( 1- \frac{1}{\Xi_{\eta_-}} \right).  \label{eq5}
\end{equation} 
We have \(0\leq \Xi_{\eta_-} \leq 1\), and \(\mbox{tr}(W) \geq 0\) as the maximally mixed state is a separable state, 
so that 
%As can be seen easily, 
RHS of inequality \eqref{eq5} is always non-positive.
%negative. 
Importantly, even when the true value of the function, $I(\rho_{AB})$, is positive, its measured value may be negative. Thus, if for a state $\rho^{\prime}_{AB}$, $I(\rho^{\prime}_{AB})$ satisfies $\frac{\text{tr}(W)}{4}\left( 1- \frac{1}{\Xi_{\eta_-}} \right) \leq I_m(\rho^{\prime}_{AB}) < 0$, then one might erroneously conclude a separable state as entangled. But one can overcome this detection loophole when the measured value is strictly less than $\frac{\text{tr}(W)}{4}(1-\frac{1}{\Xi_{\eta_-}})$. 
%\textbf{
Therefore, inequality (\ref{eq5}) provides an upper bound on the measured value of the witness function for guaranteeing entanglement even in the presence of a non-unit 
\(\Xi_{\eta_-}\).
%which will guarantee entanglement.} 

Now, if we consider the two MDI-EWs for detecting entanglement in the Werner state and genuine three-party entanglement in the noisy GHZ state, and use Eqs. \eqref{W1}, \eqref{W2}, and \eqref{eq5}, we get the following two inequalities,
\begin{eqnarray}
&&I(\rho_p)_m < \frac{1}{4}\left( 1- \frac{1}{\Xi_{\eta_-}} \right), \nonumber \\
&&I(\rho_q^{GHZ})_m < \frac{3}{8}\left( 1- \frac{1}{\Xi_{\eta_-}} \right).  \label{final}
\end{eqnarray}
They are respectively relevant for loophole-free MDI entanglement detection in the Werner state and loophole-free MDI genuine tripartite entanglement detection in the noisy GHZ state, for a particular lost event efficiency, $\Xi_{\eta_-}$.

\emph{Case II:} If we do the same calculation for $\Xi_{\eta_-}=1$ and $\Xi_{\eta_+}=\frac{\tilde{N}}{\tilde{N}+\Xi_{\epsilon_-}} \neq 1$, we get
\begin{equation}
I_m(\rho_{AB}) < \frac{\text{tr} (W)}{4} \left(1-\Xi_{\eta_+} \right). \label{final1}
\end{equation}
Here, the RHS is positive, and hence, even if the range of states that would be detected by $I_m$ is less than the states that could be detected by $I_t$, $I_m$ will not show any separable state as entangled. Hence, inefficiency due to additional count will not lead an experimenter wrongly assign a separable state as entangled while using MDI-EWs, just like for EWs. 

As we have mentioned before, the number 4 in the denominator of the right-hand-sides of the inequalities 
\eqref{eq5} and \eqref{final1} are appearing due to the fact that $a,b$ can take 4 different values, viz., 00, 01, 10, 11. 
Notice that if instead of bipartite entanglement, we consider an MDI-EW for detecting $n$-partite entanglement shared between $n$-parties, we will have $2^n$ in place of 4 and all other parts in the inequalities \eqref{eq5} and \eqref{final1} will remain unchanged.  
 %(lost event efficiency) and $\eta_+$ (additional event efficiency). Here also we will 

\begin{figure}[tb]%[ht!]
\includegraphics[width = 0.5\textwidth]{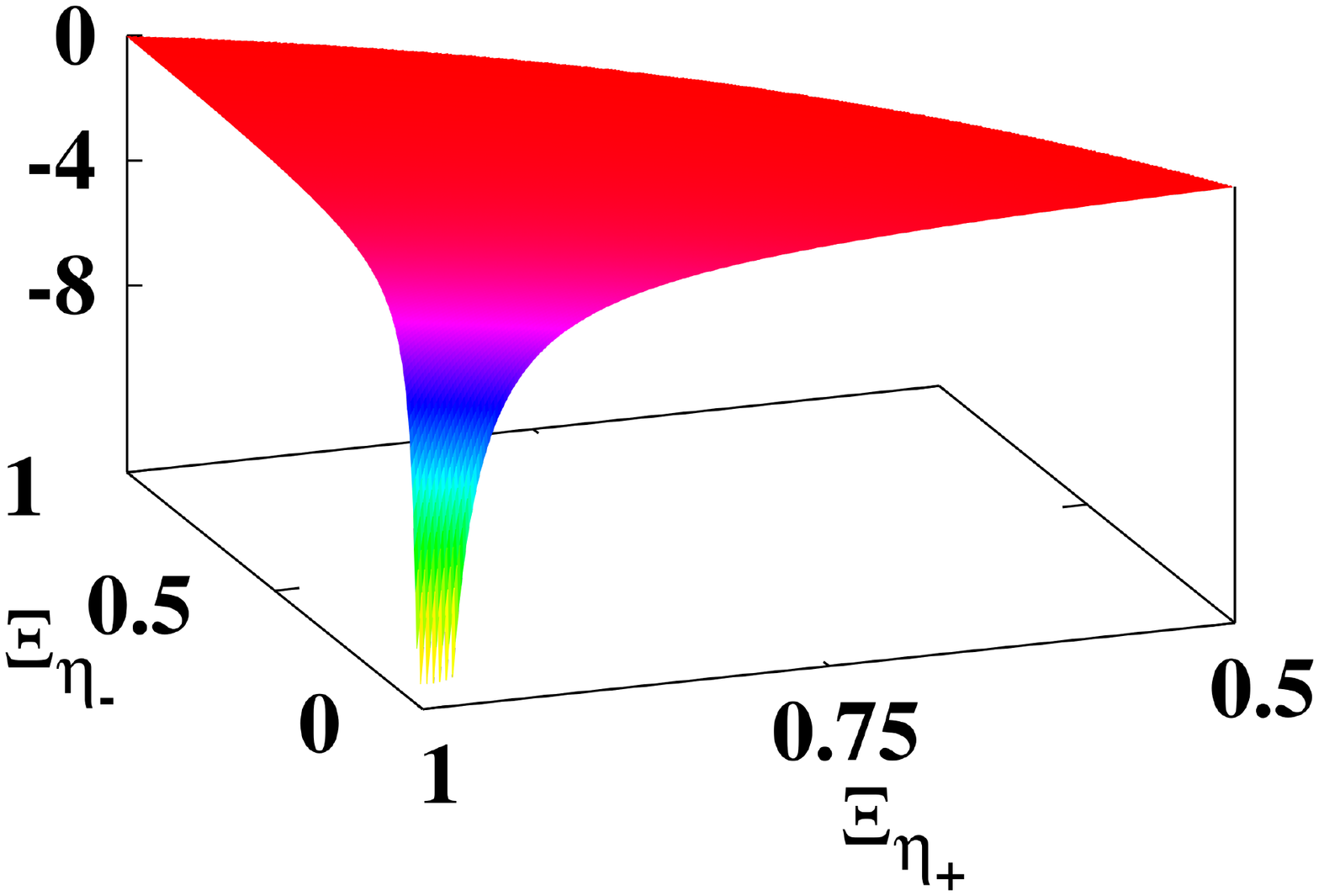} 
\caption{Beating detection loophole for measurement-device -independent entanglement witness for entangled Werner states. 
The plotted surface is such that if 
%An upper bound on 
the measured MDI-EW function, $I_m$, for an entangled Werner state 
is below it, then the experimentalist will  
%that can be used to 
be able to detect the entanglement in that Werner state in a loophole-free and MDI way.
The vertical axis represents the values of that surface, and it is 
%avoid the detection loophole, is 
plotted in the figure with respect to lost 
%event efficiency, $\Xi_{\eta_-}$, 
and additional event efficiencies, respectively $\Xi_{\eta_-}$ and $\Xi_{\eta_+}$, on the base. The surface remains negative for the region $\Xi_{\eta_-}+\frac{1}{\Xi_{\eta_+}} < 2$. All axes represent dimensionless quantities.} \label{fig.1}
\end{figure}

\emph{Case III:} We now consider the general scenario, where the additional event efficiency, $\Xi_{\eta_+}$, as well as the lost event efficiency, $\Xi_{\eta_-}$, 
are both non-ideal, i.e., they both possess non-unit values.
%both are kept arbitrary. 
A relation involving 
%upper bound on 
the measured MDI-EW function, $I_m$, which guarantees entanglement present in a bipartite state $\rho_{AB}$, is given by
\begin{equation}
I_m(\rho_{AB}) < \frac{\text{tr} (W)}{4} \left(1-\frac{1}{\Xi_{\eta_-}+\frac{1}{\Xi_{\eta_+}}-1}\right).
\label{bishop}
\end{equation}
Note that the bound on $I_m$ reaches zero, which is the bound on  MDI-EW function without any loophole in outcomes, when 
\begin{equation}
\Xi_{\eta_-}+\frac{1}{\Xi_{\eta_+}}=2.
\label{queen}
\end{equation}
It is important to know the cases when the bound on $I_m$ is negative, 
since these instances may lead to false identifications of separable states as entangled ones. 
%as then there exist chances of misleading a separable state to be entangled. 
For non-negative bounds, the loophole due to lost and additional events vanishes.
To visualize the bound, we plot, in Fig. \ref{fig.1}, the upper bound on the measured MDI-EW function, $I_m$, for entangled Werner states 
%is plotted with respect to 
with varying 
%lost event efficiency, 
$\Xi_{\eta_-}$
%, and additional event efficiency, 
and $\Xi_{\eta_+}$, in the non-trivial part of the region demarcated by Eq. (\ref{queen}). 
%It is clear that for $\Xi_{\eta_-}+\frac{1}{\Xi_{\eta_+}}\geq 2$, the bound becomes positive, and thus existence of the loophole will vanish.} 

The bound on $I_m$, given by the relation (\ref{bishop}), can easily be generalized to a genuine $n$-partite  MDI entanglement witness by  replacing 4 with $2^n$ in the denominator on the RHS.
\section{Conclusion}
%A kind of entanglement witness, known as 
A measurement-device-independent entanglement witness had been prescribed, based on the idea of a semi-quantum game, where every entangled state yields advantage over all separable states. Here we considered the problem of detection loophole in detecting entanglement in a measurement-device-independent way. We discussed three separate cases of inefficient measurement devices, viz., the case when there are only losses in the  outcomes of measurement, the case when there are only additional counts in the same, and finally the scenario where both types of the inefficiencies occur.  We found that in the case of additional events, inefficiency of the measurement device cannot lead to false positives, i.e., the measurement-device-independent entanglement witness will not show any separable state as entangled. However, we showed that in the the case of lossy detectors as well 
as in the case when both types of inefficiencies are present, measurement-device-independent entanglement witnesses can erroneously exhibit a separable state as entangled. To avoid such misjudgments, 
%overcome such errors, 
we derived an
inequality in each case, which depend on the  efficiencies of the detectors involved. If a measured value of the measurement-device-independent entanglement witness satisfies the relevant inequality, for a particular set of values of the detector efficiencies, then one will reach the unambiguous conclusion of presence of entanglement in the shared quantum state under consideration. 
We have exemplified the results by using the Werner and noisy Greenberger-Horne-Zeilinger states.

%We have also compared the bounds on the measured values of the witness functions, which guarantee entanglement, in both the scenarios,  viz., measurement-device-dependent and measurement-device-independent entanglement witnesses, with non-ideal measurement devices. 
%By comparing the two results we gave a condition for choosing more suitable one between EW and MDI-EW for detecting entanglement in a measurement device where detection loophole is present but the measurement performed. 

\label{sec5}

\end{document}